\documentclass[a4paper,11pt]{article}
\usepackage{pos}

\title{UCIRC2: EUSO-SPB2's Infrared Cloud Monitor}

\ShortTitle{UCIRC2}

\author*[a]{Rebecca Diesing}
\author[a]{Alexa Bukowski}
\author[a]{Noah Friedlander}
\author[a]{Alex Miller}
\author[a]{Stephan Meyer}
\author[a]{Angela V. Olinto}

\affiliation[a]{Department of Astronomy \& Astrophysics, KICP, EFI, The University of Chicago, IL 60637, USA\\}
\forColl{JEM-EUSO} % W/O "Collaboration"

\emailAdd{rrdiesing@uchicago.edu}

\abstract{The second generation of the Extreme Universe Space Observatory on a Super Pressure Balloon (EUSO-SPB2) is a balloon instrument for the detection of ultra high energy cosmic rays (UHECRs) with energies above 1 EeV and very high energy neutrinos with energies above 10 PeV. EUSO-SPB2 consists of two telescopes: a fluorescence telescope pointed downward for the detection of UHECRs and a Cherenkov telescope pointed towards the limb for the detection of tau lepton-induced showers produced by up-going tau neutrinos and background signals below the limb. Clouds inside the field of view of these telescopes reduce EUSO-SPB2’s geometric aperture, in particular that of the fluorescence telescope. For this reason, cloud coverage and cloud-top altitude within the field of view of the fluorescence telescope must be monitored throughout data-taking. The University of Chicago Infrared Camera (UCIRC2) will monitor these clouds using two infrared cameras with response centered at wavelengths 10 and 12 microns. By capturing images at wavelengths spanning the cloud thermal emission peak, UCIRC2 will measure cloud color-temperatures and thus cloud-top altitudes. In this contribution, we provide an overview of UCIRC2, including an update on its construction and a discussion of the techniques used to calibrate the instrument.}

\FullConference{37th International Cosmic Ray Conference, ICRC2021\\
		12 - 23 July, 2021\\
		Berlin, Germany}

\begin{document}
\maketitle

\section{Introduction}

Ultra High Energy Cosmic Rays (UHECRs), cosmic rays (CRs) with energy above $10^{18}$ eV, are currently detected with Cherenkov water tanks and up-looking fluorescence detectors in observatories such as the Pierre Auger Observatory \cite{Auger} in Argentina and the Telescope Array \cite{TA} in Utah. In particular, UHECRs can be detected via a characteristic particle shower, called an Extensive Air-Shower (EAS), that occurs when an UHECR interacts with Earth's atmosphere. This EAS produces fluorescence of atmospheric nitrogen molecules, detectable in the 300-400 nm spectral band, as well as optical Cherenkov light. Because UHECRs are rare, ($<$ 1 per km${^2}$ per century close to $10^{20}$ eV), extremely large detector volumes are required to enable charged-particle astronomy. One way to increase detector volume is to observe the atmosphere from above. This is what the second generation of the Extreme Universe Space Observatory on a Super Pressure Balloon (EUSO-SPB2) will do.

\begin{figure}
\begin{center}
\includegraphics[width=0.95\textwidth, clip=true,trim= 0 0 20 0]{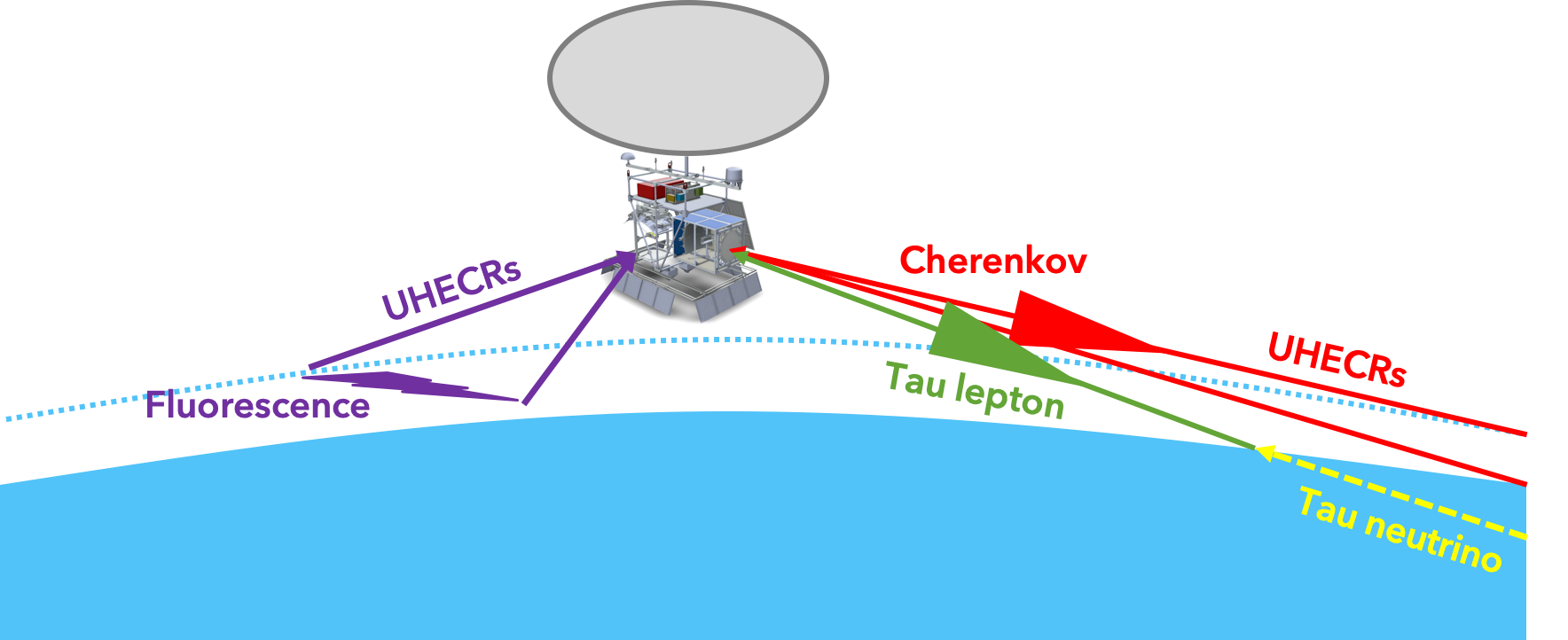}   
\end{center}
\caption{EUSO-SPB2's three detection modes: fluorescence from UHECRs (purple), Cherenkov from UHECRs (red), and Cherenkov from CNs (green).}
\label{fig:EUSOSPB2}
\end{figure}

A pathfinder to the satellite mission POEMMA (Probe of Extreme Multi-Messenger Astrophysics), EUSO-SPB2 will detect UHECRs via two complementary techniques: looking down upon the atmosphere with a fluorescence telescope and looking towards the limb of the Earth to observe the Cherenkov signals produced by UHECRs above the limb. EUSO-SPB2 will also search for the signatures of neutrinos above $10^{16}$ eV via the Cherenkov light from upward going tau leptons produced when a tau neutrino interacts near the surface of the Earth (see Figure \ref{fig:EUSOSPB2}) \cite{EUSOSPB2}. 

The presence of high clouds within the detectors' field of view (FoV) can significantly reduce the UHECR event detection rate and event energy calibration. Namely, it is possible for some of the EAS signal to occur behind high clouds. Determining the exposure of EUSO-SPB2 to UHECRs requires knowledge of the effective detector volume, i.e., the volume of atmosphere within the field of view, above the clouds. Thus, EUSO-SPB2 requires continuous information about cloud coverage and height within the detectors' FoV. This is the responsibility of the second generation of the University of Chicago Infrared Camera (UCIRC2). 

The design and calibration of UCIRC2 improves upon that of UCIRC1, which flew onboard EUSO-SPB1. More information on UCIRC1 can be found in \cite{UCIRC1}.

\section{Method}

\begin{figure}
    \centering
    \includegraphics[width=0.95\textwidth]{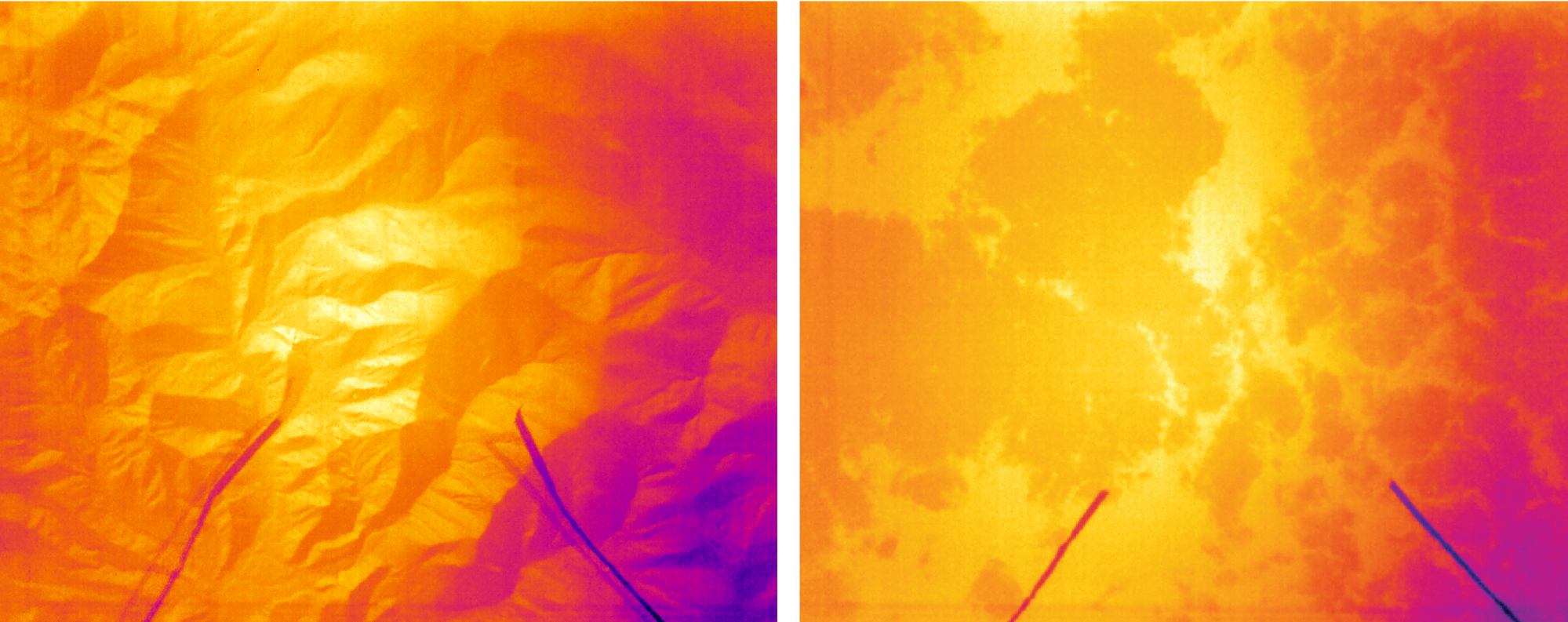}
    \caption{Uncalibrated images of mountains (left) and clouds (right) captured by UCIRC1, which flew on EUSO-SPB1 in 2017. Even without calibration, cloud coverage can be easily determined.}
    \label{fig:SamplePic}
\end{figure}

When EUSO-SPB2 is in observing (night) mode, IR images of the environmental conditions in and around the effective UHECR detection area are captured by UCIRC2 every 120 seconds. These images can be used to collect information about cloud coverage and altitude (cloud top height, CTH) within the field of view of the UHECR detectors (see Figure \ref{fig:SamplePic}).

Because the clouds are at the temperature of the air, CTH can be inferred from cloud temperature, $T_{\rm c}$ which can be estimated using two brightness temperatures in bands near in wavelength to the cloud blackbody peak. More specifically, UCIRC2's two IR cameras observe at wavelengths of 10$\mu$m and one at 12$\mu$m. A calibrated image in a single frequency band can be used to determine the temperature of an object of known emissivity ($\epsilon$), but cloud emissivity is highly variable and significantly less than 1. Thus, a multifrequency observation is required to break the degeneracy between $\epsilon$ and $T_{\rm c}$. For a single layer of clouds above an ocean of known surface temperature and reflectivity (and thus power, $P_{\rm E}$), one can estimate power on the detector, $P_{\rm tot}$ as,

\begin{equation}
    P_{\rm tot} = \epsilon P_{c}+(1-\epsilon)P_{E}
\end{equation}

Here, $P_{c}$ is the power of the cloud, from which $T_{c}$ and thus CTH, can be inferred. Other methods for reconstructing CTH can be found in \cite{Mario}.

\section{Design}

\subsection{IR Cameras}

UCIRC2 will be outfitted with two $640\times480$ pixel Teledyne DALSA Calibir GXF uncooled IR cameras with 14mm lenses, focused at infinity. The cameras have a $42^\circ \times 32^\circ$ FoV, chosen to be somewhat larger than that of EUSO-SPB2's fluorescence telescope. When the payload is in ``night mode'', which occurs when the atmosphere is dark enough to allow for proper functioning of photodetection modules (PDMs), UCIRC takes a pair of pictures every two minutes. The wide field of view of the IR cameras  makes it possible to extrapolate the cloud conditions in the section of the atmosphere swept out by the PDM field of view in the time between pictures. 

The native spectral response of the cameras is 8 to 14 microns, but each camera is fitted with a filter to facilitate the radiative CTH reconstruction. One of the cameras is fitted with a SPECTROGON bandpass light filter which transmits wavelengths between 11.5 and 12.9$\mu$m (denoted 12$\mu$m). The other with an Edmund Optics bandpass light filter that transmits wavelengths between 9.6 and 11.6$\mu$m (denoted 10$\mu$m). These bands are spaced to obtain brightness temperature data that facilitates both the Blackbody Power Ratio CTH reconstrction method and the Radiative Transfer Equation CTH reconstruction method. 

The cameras are powered via a 12V connection and communicate via Gigabit Ethernet with a single-board, industrial-grade CPU that can operate at temperatures between -40C and 85C. 

\subsection{Software}

The cameras can be operated via CamExpert, a graphic user interface (GUI) which generates images from IR camera output. In order to automatically capture and store lossless images without a GUI, we will develop software using Sapera LT, a development kit (SDK) that accompanies Calibir-series cameras. This software will enable the acquisition of raw ADC output from each microbolometer pixel as well as the temperature of the sensor array, which affects the cameras' response. In order to improve the signal to noise ratio of the resulting data, a burst of images will be captured every two minutes and added together. This ``sum image" is then compressed using bzip2, such that it requires $\lesssim 250$kB of storage on the UCIRC2 CPU. Since two sum images (one per camera) are captured every two minutes, UCIRC2 will generate approximately 75MB of data per night, assuming astronomical night lasts $\sim 5$ hours.

\subsection{Environment Control}

\begin{figure}
\begin{center}
\includegraphics [width=.95\textwidth]{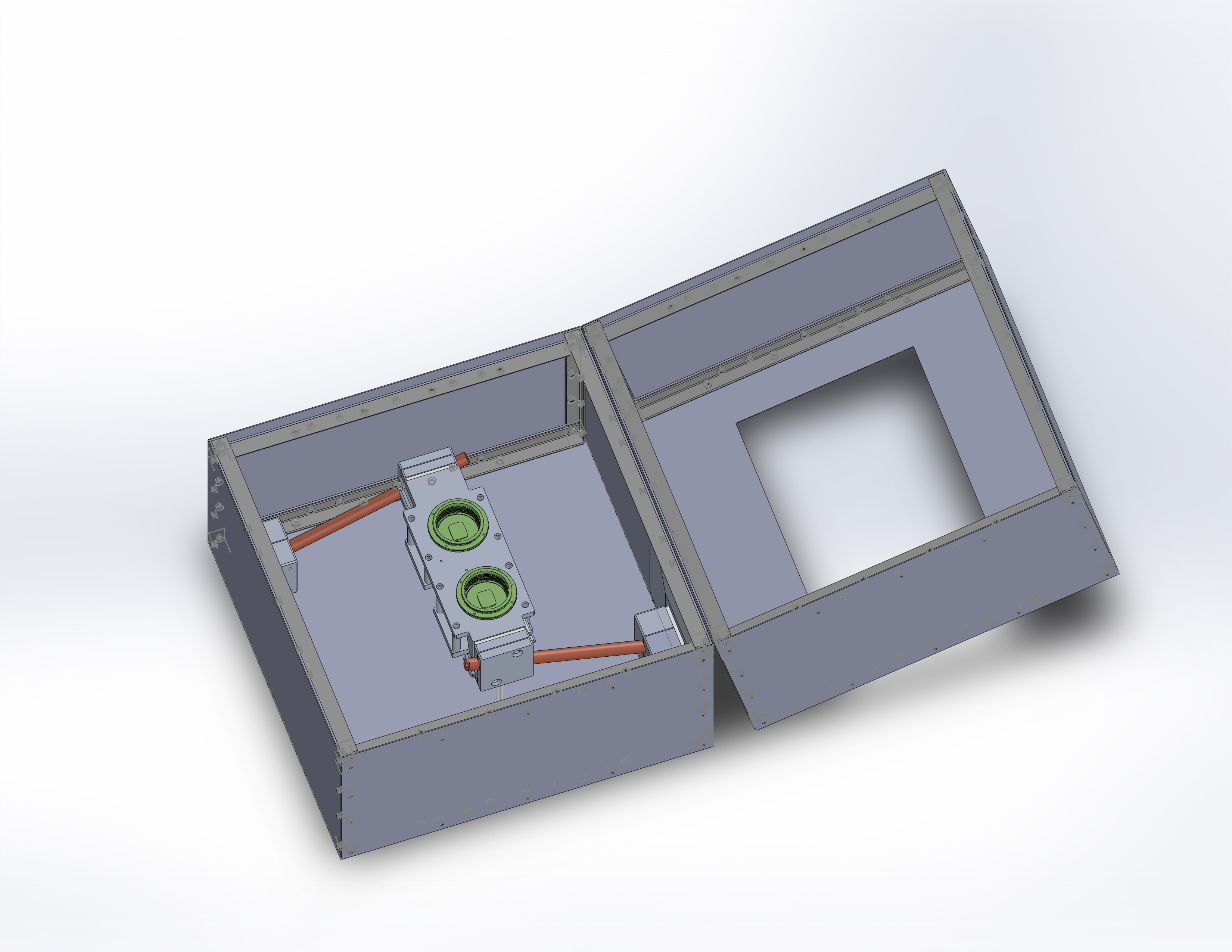}
\caption{Rendering of UCIRC2, with cameras (green circles) pointed toward the viewer and aluminum box open. 10 and 12 micron filters will be mounted in front of each camera. A system of peltier coolers (white rectangular prisms), heat pipes (copper-colored tubes), and resistive heaters (not shown) will maintain a steady camera temperature. The entire system will be enclosed in an aluminum box (grey structure) and coated with emissive paint (not shown).}
\label{fig:schematic}
\end{center}
\end{figure}

\begin{figure}
    \centering
    \includegraphics[width=0.95\textwidth]{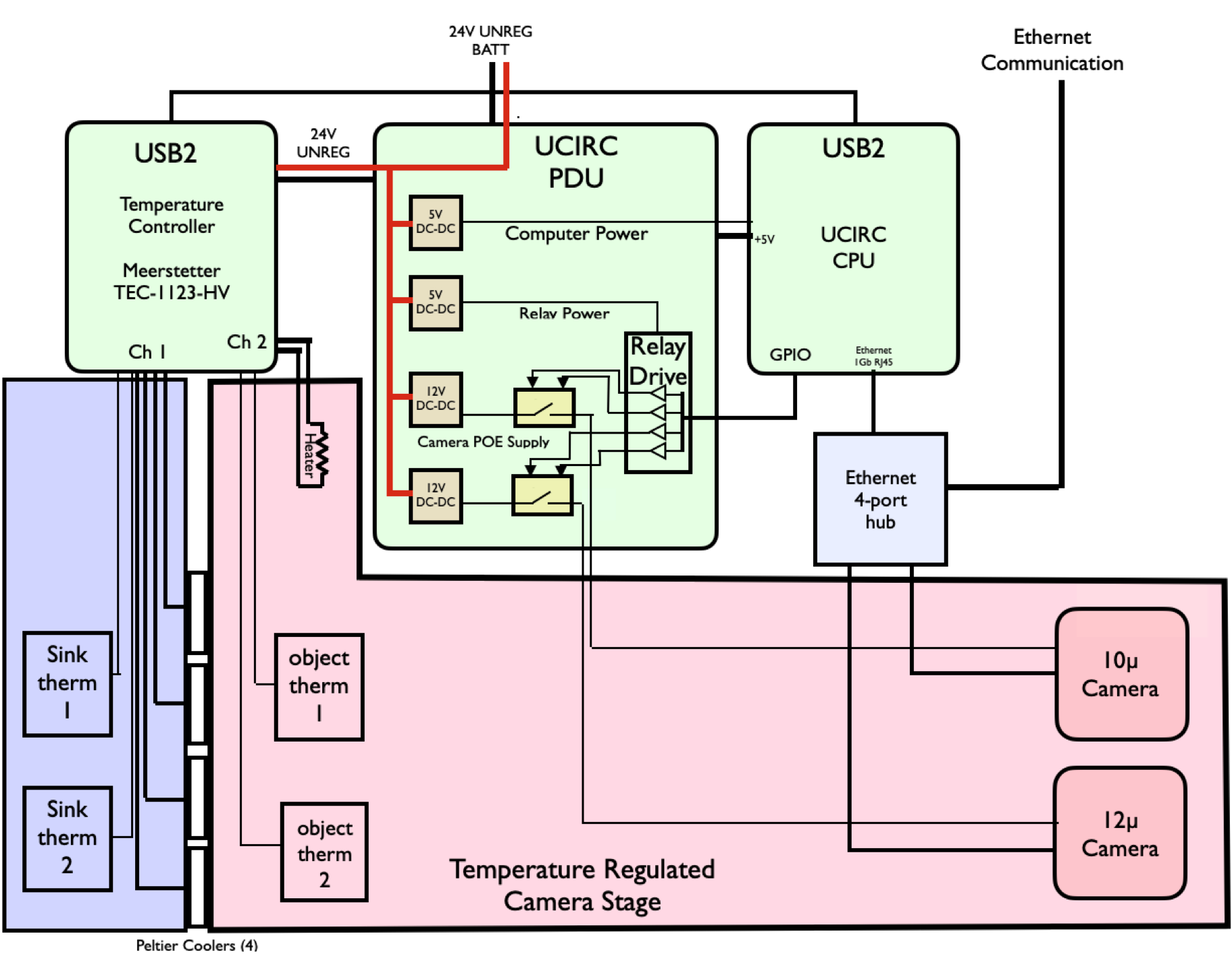}
    \caption{A block diagram of the UCIRC2 electronics and temperature management system. The two cameras communicate with the UCIRC2 CPU via ethernet using a two port hub. The CPU initializes image acquisition, compresses images, and stores them. The cameras are mounted on a temperature stabilized platform controlled via a thermal controller using Peltier coolers and a resistive heater. Unregulated 24V battery power is converted to 5V for the CPU and temperature controller and 12V for the cameras. Power to the cameras can be individually be switched on and off by the CPU. The electronics are passively cooled via heat pipes.}
    \label{fig:thermal}
\end{figure}

UCIRC2 is designed to operate in a high altitude ($\approx 33$km) environment during both daytime, when ambient temperatures reaches approximately 40C, and nighttime, when ambient temperatures reach approximately -40C. Temperature management is therefore a central design concern. In particular, the camera response is temperature dependent, meaning that camera temperature must be held approximately constant during operation (night mode). To maintain a stable temperature, the two cameras are housed in a $300$mm$\times$300mm$\times300$mm aluminum box coated with high emissivity flat white paint. This box is hinged and can be easily opened and closed to allow access to the cameras and electronics. An active temperature management system consisting of heaters, Peltier coolers, heat pipes, and thermometers enables precise temperature monitoring and control (see Figures \ref{fig:schematic} and \ref{fig:thermal}).

The active heating and cooling system is controlled by a two-channel Meerstetter Engineering HV-1123 thermoelectric cooling and heating controller (TEC). Channel 1 of the TEC drives current to four Laird Technologies 56995-501 30mm$\times$30mm Peltier coolers. The second channel of the TEC is connected to a 10 $\Omega$ Vishay Dale resistive heater. The resistive heater is the primary mechanism for heating the camera stage and is used in combination with the Channel 1 Peltier system when heating is needed. The active temperature control system is designed to be most effective at heating because, in general, UCIRC2 will be actively collecting data during the nighttime, when the environment is cold. 

The temperature regulated camera stage is a machined aluminum plate, called the main stage, to which both IR cameras are thermally coupled. This stage is connected with good thermal contact to two Peltier coolers, which connect via two Enertron sintered powder wick copper heat pipes leading to two additional Peltier units at the outside panels. These heat pipes transfer the heat pumped by the Peltier units between the camera stage and the aluminum side walls. 

A second thermal stage supports the electronics boards (the CPU, the TEC, the USB hub and the power distribution board). This stage is cooled passively by heat pipes.

The TEC-1123-HV uses four temperature sensors. Two NTC thermistors with 5k$\Omega$ resistance at 20C measure the heat sink temperature, where two of the Peltier coolers contact the side walls. The other two sensors, Platinum Resistance Thermometers (PRTs) operated with a four wire readout, measure the temperature of the center and the edge of the camera thermal stage. The temperature measurement of the center of the camera thermal stage controls the TEC Proportional Integral Derivative controller (PID) in order to keep the main thermal stage at a constant, settable temperature. The second sensor is run as a monitor but does not control the PID loop. The set point temperature for the cameras can be modified by telemetery command, with daytime and nighttime operating temperatures chosen to minimize power consumption. 

\section{Testing and Calibration}

\begin{figure}[ht]
    \centering
    \includegraphics[width=1.0\textwidth]{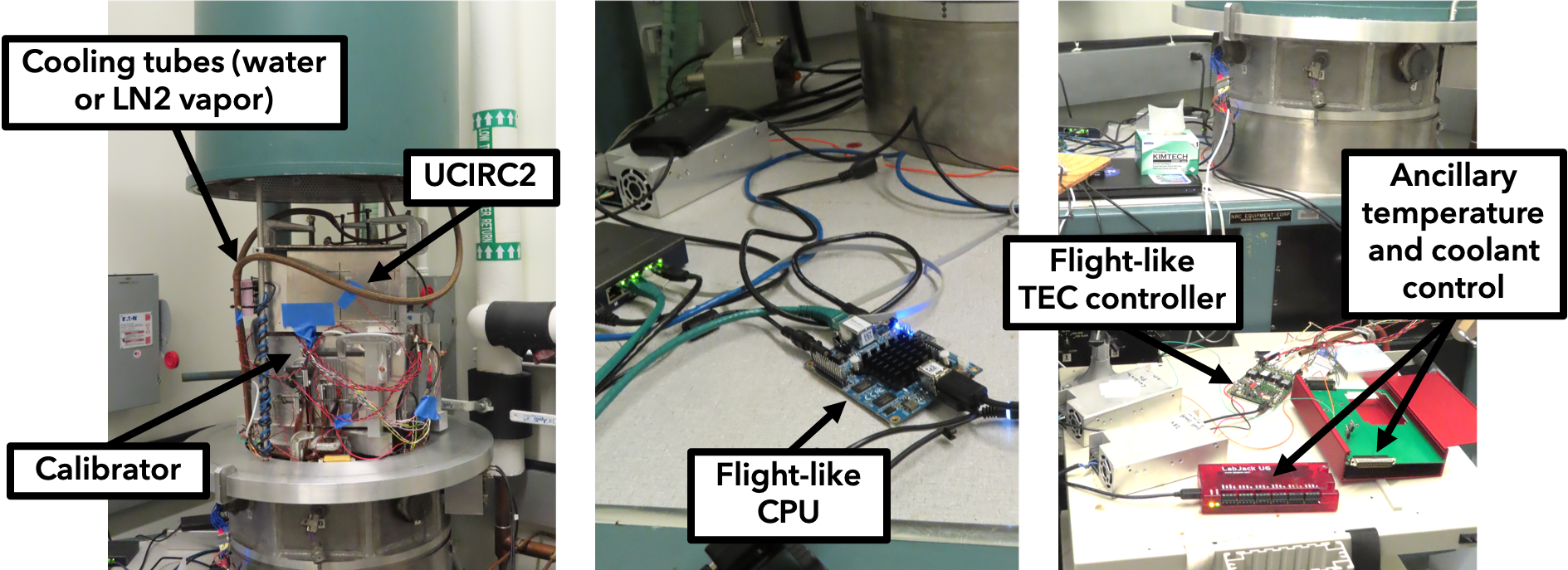}
    \caption{Photographs of the UCIRC2 calibration setup, in which a prototype of UCIRC2 is mounted above a calibrator inside a thermovac chamber (pictured in its open position for illustrative purposes). This calibrator consists of an aluminum box housing a temperature-controlled blackbody target. As in UCIRC2, the temperature of the target is controlled by a system of peltier coolers, heat pipes, and resistive heaters.}
    \label{fig:calibrator}
\end{figure}

To replicate the expected flight environment, a protype of UCIRC2 has been tested in a thermovac chamber pumped down to 0.3 mbar and a shroud cooled with liquid nitrogen vapor. The temperature management system will be tested over all possible environmental temperatures to ensure that the cameras can be maintained within their operating temperature range. 

To calibrate the cameras, UCIRC2 is positioned above a calibration target consisting of a highly emissive, temperature-controlled material (see Figure \ref{fig:calibrator}). By taking of images of the calibration target at multiple temperatures, this target can be used to perform a pixel by pixel calibration of each camera. Because the cameras' thermal response depends on their temperature, calibration images will also be taken at multiple camera temperatures. Before flight, the camera response will be tested under vacuum over all potential environmental and target temperatures.

The prototype of UCIRC2 pictured in Figure \ref{fig:calibrator} uses our original choice of camera, the $640\times480$ pixel Sierra Olympic Viento G in place of the Teledyne DALSA Calibir GXF. We completed calibration of these cameras over a wide range of camera and target temperatures, both with and without liquid nitrogen vapor cooling. This calibration revealed the cameras to be noisier than advertised and extremely sensitive to their own temperature, with their response varying by $\gtrsim 300$ counts per degree Kelvin (ct/K) compared with a response of only $\simeq 13$ ct/K to variations in calibrator temperature. Given the absence of readable thermometers on the Viento G's sensor array, these sensitivities would prohibit meaningful measurements of CTH. Thus, our calibration results prompted the acquisition of the Calibir GXF, which is explicitly designed for thermography and has a readable thermometer on its sensor array. Calibration of the Calibir GXF will proceed using the techniques and insights developed during the calibration of its predecessor.

\section{Acknowledgements}
UCIRC2 is supported by NASA Grant 80NSSC18K0246 and acknowledges previous work from the UCIRC1 team and the JEM-EUSO collaboration.

\clearpage
\section*{Full Author List: The \Coll\ Collaboration}

\begin{sloppypar}
{\small \noindent 
G.~Abdellaoui$^{ah}$, 
S.~Abe$^{fq}$, 
J.H.~Adams Jr.$^{pd}$, 
D.~Allard$^{cb}$, 
G.~Alonso$^{md}$, 
L.~Anchordoqui$^{pe}$,
A.~Anzalone$^{eh,ed}$, 
E.~Arnone$^{ek,el}$,
K.~Asano$^{fe}$,
R.~Attallah$^{ac}$
% ...
}
\end{sloppypar}
\vspace*{.3cm}

%%\newpage
{ \footnotesize
\noindent
% Algeria (Dezember 2013) - 7 institutes
% ...
$^{ac}$ LPR at Dept. of Physics, Faculty of Sciences, University Badji Mokhtar, Annaba, Algeria \\
% ...
$^{ah}$ Telecom Lab., Faculty of Technology, University Abou Bekr Belkaid, Tlemcen, Algeria \\
% Bulgaria ready (02042012)  - 1 institutes 
% ...
$^{cb}$ Universit\'e de Paris, CNRS, AstroParticule et Cosmologie, F-75013 Paris, France\\
% ...
$^{ed}$ Istituto Nazionale di Fisica Nucleare - Sezione di Catania, Italy\\
% ...
$^{eh}$ INAF - Istituto di Astrofisica Spaziale e Fisica Cosmica di Palermo, Italy\\
% ...
$^{ek}$ Istituto Nazionale di Fisica Nucleare - Sezione di Torino, Italy\\
$^{el}$ Dipartimento di Fisica, Universita' di Torino, Italy\\
% ...
$^{fe}$ Institute for Cosmic Ray Research, University of Tokyo, Kashiwa, Japan\\ 
% ...
$^{fq}$ Nihon University Chiyoda, Tokyo, Japan\\ 
% ...
$^{md}$ Universidad Polit\'ecnia de madrid (UPM), Madrid, Spain\\
% ...
$^{pd}$ University of Alabama in Huntsville, Huntsville, AL; USA\\
$^{pe}$ Lehman College, City University of New York (CUNY), NY, USA\\
% ...
%16 Leerzeilen in Affils.
}

\end{document}